\documentclass[a4paper,11pt]{article}
\usepackage{pos}
\usepackage{siunitx}
\usepackage{subcaption} % to have multiple plots side by side

\usepackage{float}
\usepackage{url}
\usepackage{hyperref}
\usepackage[backend=biber,maxcitenames=1,style=science]{biblatex}

\newcommand{\simtel}{sim\_telarray}

\newcommand{\hess}{H.E.S.S.}

\title{Validating Monte Carlo simulations for an analysis chain in H.E.S.S.}
\author*[a]{F. Leuschner}
\author[b]{J. Schäfer}
\author[c]{S. Steinmassl}
\author[d]{T. L. Holch}
\author[c]{K. Bernlöhr}
\author[b]{S. Funk}
\author[c]{J. Hinton}
\author[d]{S. Ohm}
\author[a]{G. Pühlhofer}

\affiliation[a]{Institut für Astronomie und Astrophysik, Eberhard Karls Universität Tübingen, %\\
  D 72076 Tübingen, Germany}

\affiliation[b]{Erlangen Centre for Astroparticle Physics, Friedrich-Alexander-Universität Erlangen-Nürnberg, %\\
D 91058 Erlangen, Germany}

\affiliation[c]{Max-Planck-Institut für Kernphysik (MPIK), %\\
D 69029 Heidelberg, Germany}

\affiliation[d]{Deutsches Elektronen-Synchrotron (DESY), %\\
D 15738 Zeuthen, Germany}

\emailAdd{fabian.leuschner@astro.uni-tuebingen.de}
\emailAdd{johannes.schaefer@fau.de}
\emailAdd{simon.steinmassl@mpi-hd.mpg.de}
\emailAdd{tim.lukas.holch@desy.de}
\emailAdd{Konrad.Bernloehr@mpi-hd.mpg.de}
\emailAdd{s.funk@fau.de}
\emailAdd{jim.hinton@mpi-hd.mpg.de}
\emailAdd{stefan.ohm@desy.de}
\emailAdd{Gerd.Puehlhofer@astro.uni-tuebingen.de}

\abstract{Imaging Air Cherenkov Telescopes (IACTs) detect very high energetic (VHE) gamma rays. They observe the Cherenkov light emitted in electromagnetic shower cascades that gamma rays induce in the atmosphere. A precise reconstruction of the primary photon’s energy and the source flux depends heavily on accurate Monte Carlo (MC) simulations of the shower propagation and the detector response, and therefore also on adequate assumptions about the atmosphere at the site and time of a measurement.\\
Here, we present the results of an extensive validation of the MC simulations for an analysis chain of the H.E.S.S. experiment with special focus on the recently installed FlashCam camera on the large \SI{28}{\m} telescope. One goal of this work was to create a flexible and easy-to-use framework to facilitate the detailed validation of MC simulations also for past and future phases of the H.E.S.S. experiment.\\
Guided by the underlying physics, the detector simulation and the atmospheric transmission profiles were gradually improved until low level parameters such as cosmic ray (CR) trigger rates matched within a few percent between simulations and observational data. This led to instrument response functions (IRFs) with which the analysis of current H.E.S.S. data can ultimately be carried out within percent accuracy, substantially improving earlier simulations.}

\FullConference{%
  7th Heidelberg International Symposium on High-Energy Gamma-Ray Astronomy (Gamma2022)\\
  4-8 July 2022\\
  Barcelona, Spain\\}

\begin{document}
\maketitle

\section{Introduction}
MC simulations are crucial for the analysis of data taken by IACTs including the creation of IRFs for these instruments. A prominent toolkit for IACT array simulations is \simtel\ \cite{simtel2008} in connection with CORSIKA \cite{CorsikaKIT}. 
This combination is used for MC simulations within one analysis chain of the \hess\ experiment and also for studies of the upcoming Cherenkov Telescope Array (CTA).
In this work, we present the validation of simulations for an analysis chain of the \hess\ array. It consists of CT5, a large central telescope (\SI{28}{m} mirror diameter), surrounded by four small telescopes CT1-4 (\SI{12}{m} mirror diameter).
Focus is put on the FlashCam camera that was installed at the end of 2019 on CT5 \cite{ICRC21FCPerf}, \cite{ICRC21FCTechnical}. The goal is to reach consistency between MC simulations and observational data up to the DL3\footnote{For more information see \url{https://doi.org/10.3390/universe7100374}.} data level.
Repeatability is another important goal of this work which is achieved through a Python code base. 
Within these proceedings, we report on the various steps of the validation, learned lessons and reached improvements.

\section{Basic MC and single telescope validation}
\subsection{Basic MC checks}
In a first step, single telescope simulations, using a laser with a fixed photon intensity as simulated light source, were used to check the consistency of the charge integration algorithms and the photo electron definition. Thereby, consistency within a few percent between simulations and measurements could be established. \\
% Stuff about optics and PSF
To validate the optical Point Spread Function (PSF) adopted in simulations for each telescope, we use the included ray-tracing feature of \simtel\ which calculates the path of photons through the
optical system. The response of the system is evaluated using parallel light as an input. The telescope is pointed at an infinitely distant star at zenith. The off-axis response is probed by introducing a pointing shift between the telescope and the simulated star. 
An intensity map is created from the photon positions on the camera-lid and a circle is inscribed around the centre of gravity. The radius of this circle is increased until 80\,\% of the total signal is contained within. This 80\,\% containment radius is further referred to as the PSF$_{80}$ of the telescope.\\
The PSF$_{80}$ is simulated for different telescope elevations $\Theta$ and compared to measurements obtained within the same hardware phase\footnote{That is a period of time in which the hardware configuration and instrument response is considered constant. A new phase starts e.g.\ when hardware is replaced, mirrors are cleaned, software settings are changed, etc.}. The simulated PSF$_{80}$ in \simtel\ is evaluated using the following equation: 
\begin{equation}
    \text{PSF}_{80}(\Theta)=\sqrt{ \text{R}_\text{min}^2 + \text{d}_1^2 \cdot (\sin(\Theta)-\sin(\Theta_0))^2+ \text{d}_2^2 \cdot (\cos(\Theta)-\cos(\Theta_0))^2}
    \label{eq:r80_ct5}
\end{equation}

\begin{figure}[H]
	\centering
	\subcaptionbox{CT1}%
	[0.45\textwidth]{	\includegraphics[width=0.47\textwidth]{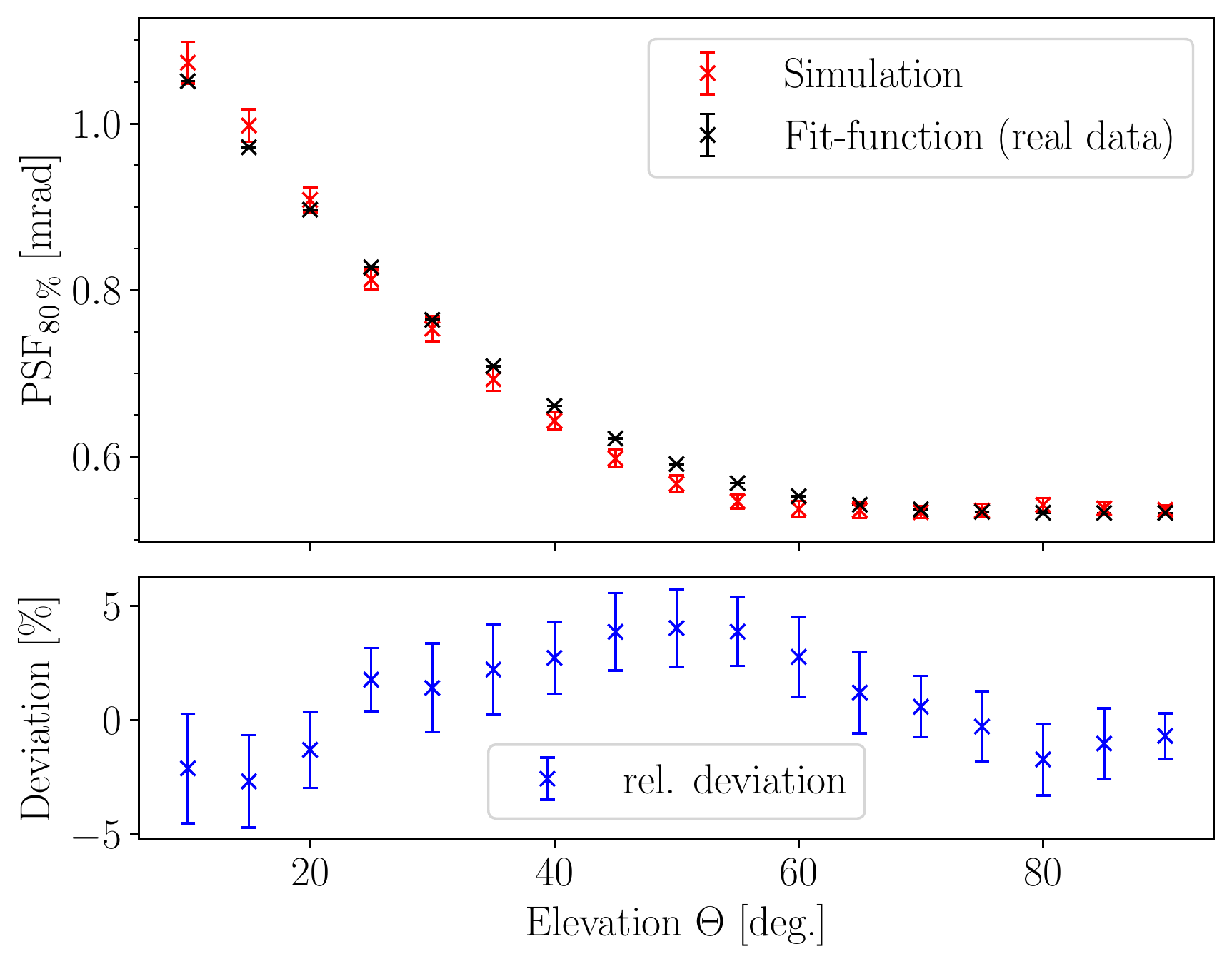}	}
	\quad
	\subcaptionbox{CT5}%
	[0.45\textwidth]{	\includegraphics[width=0.47\textwidth]{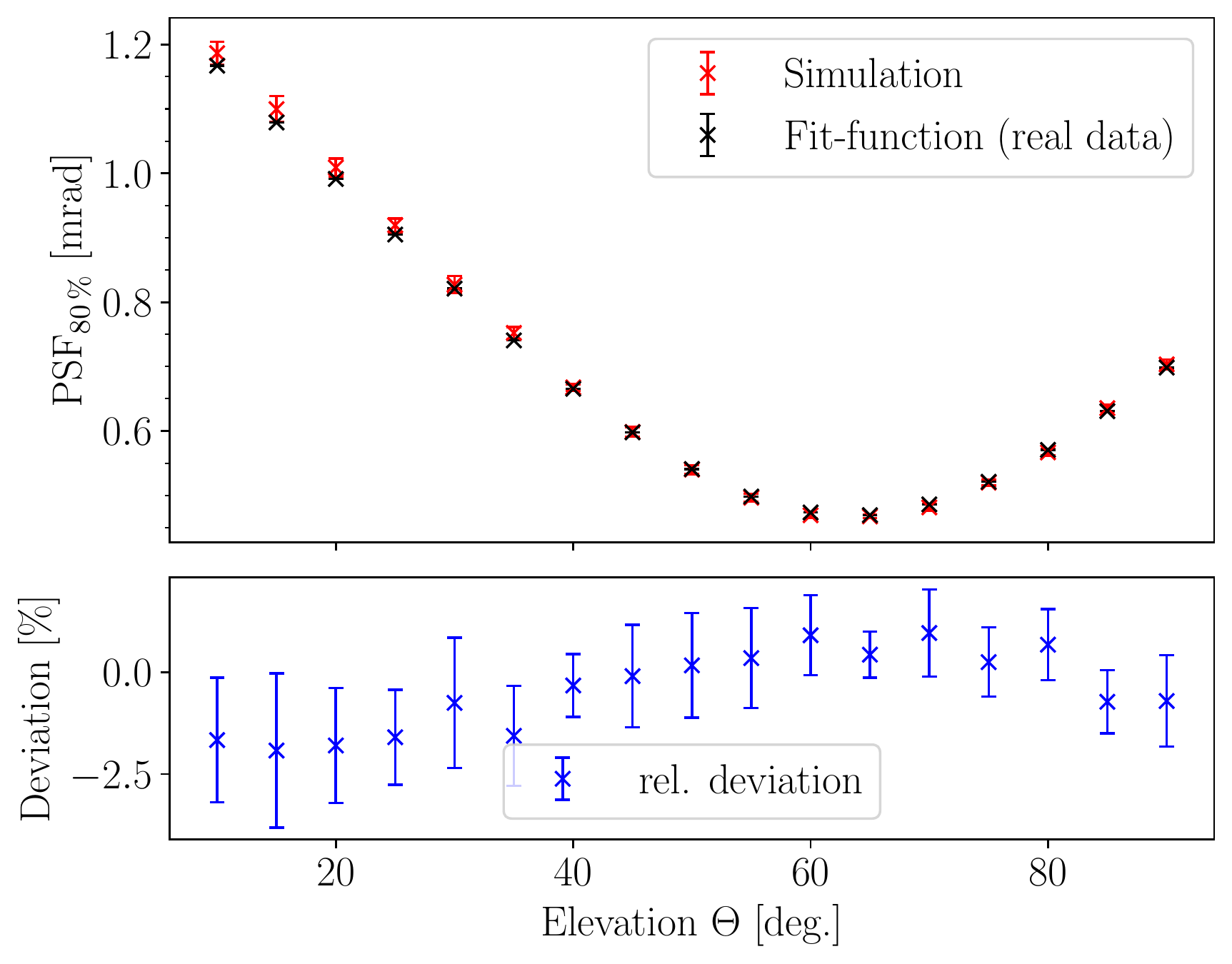}	}
	
	\caption{Optical PSF$_{80}$ for CT1 and CT5 vs. elevation based on \simtel\ ray-tracing compared to real data. A fit was made to real measurements and the resulting fit function evaluated at different elevations following \cite{CornilsPSF}. In the simulation datapoints were directly evaluated without a fitting procedure. }
	\label{Fig:PSF_80}
\end{figure}

Here, $\text{R}_\text{min}$, $\text{d}_1$, $\text{d}_2$ and $\Theta_0$ are free parameters that can be specified in the simulation configuration. These are readjusted until the overall deviation between simulated and measured PSF$_{80}$ is minimized. With the new telescope dependent parameters the PSF$_{80}$ deviation is reduced to $<5\,\% \approx 0.4\,\text{mm}$ in the focal plane (compare \autoref{Fig:PSF_80}) which is deemed acceptable.

\subsection{Single telescope consistency}
\begin{figure}[H]
	\centering
	\subcaptionbox{CT1 stereo participation trigger rate}%
	[0.45\textwidth]{\includegraphics[width=0.48\textwidth,trim=0 0 0 0, clip]{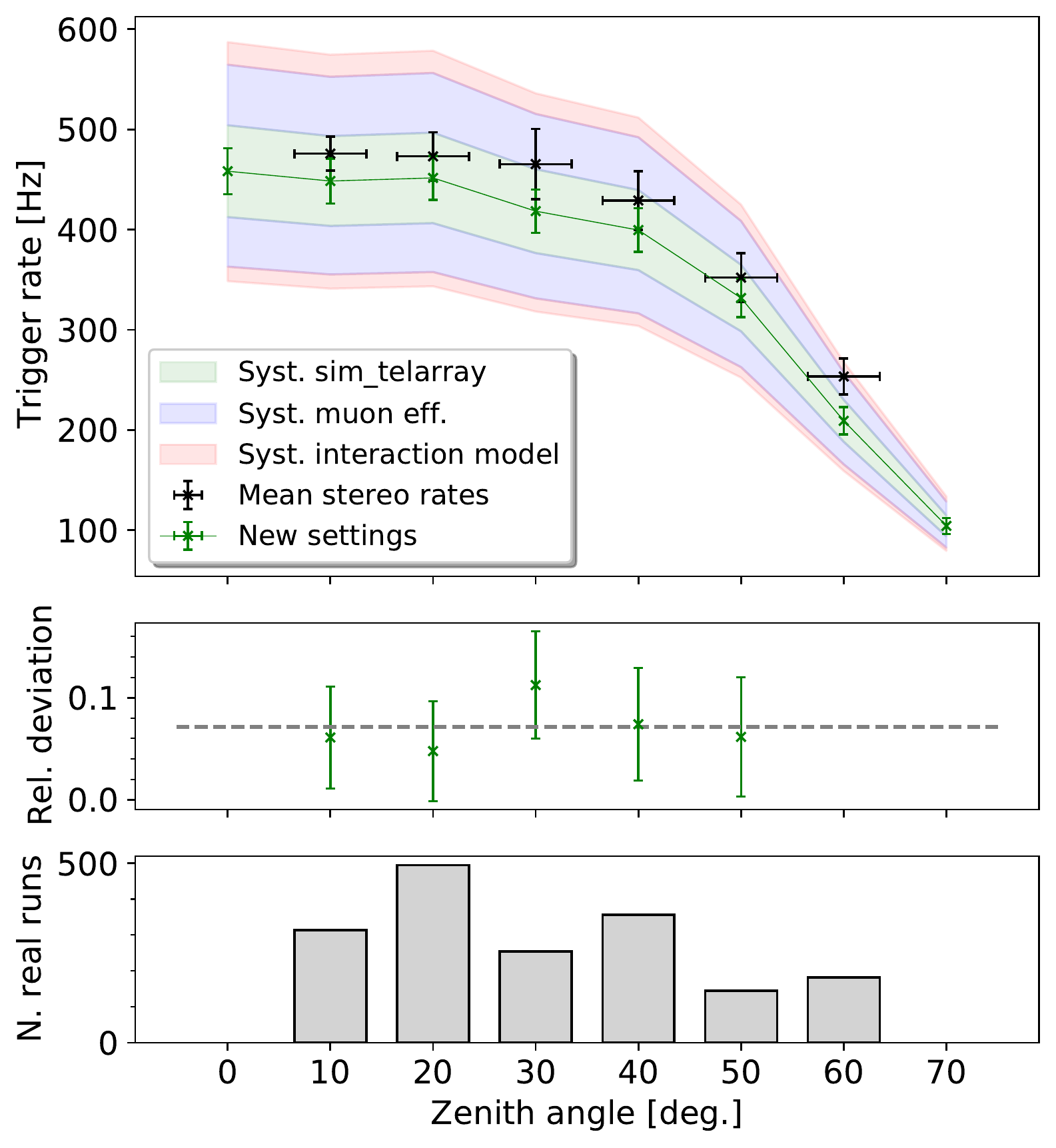}	}
	\quad
	\quad
	\subcaptionbox{CT5 mono trigger rate}%
	[0.45\textwidth]{\includegraphics[width=0.48\textwidth,trim=0 0 0 0, clip]{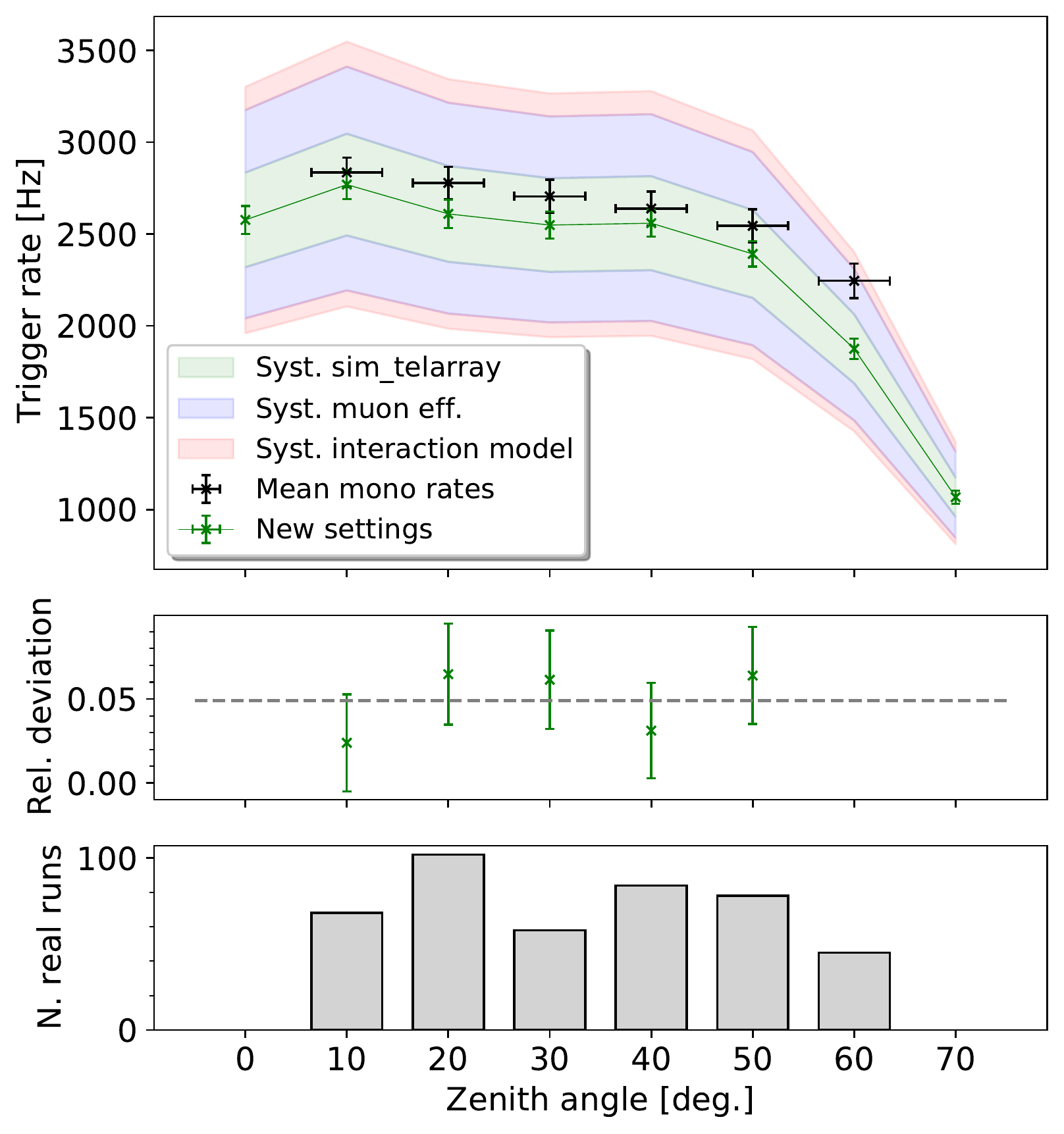}	}

	%\caption{Simulated and real zenith-dependant stereo participation and mono trigger rate for CT1 and CT5 respectively with uncertainty bands highlighting the different investigated systematical uncertainties as discussed in the main text.}
        \caption{Simulated and real zenith-dependent trigger rates for CT1 (stereo) and CT5 (mono) with uncertainty bands highlighting the different investigated systematic uncertainties as discussed in the main text.}
	\label{Fig:final_trigger_rates}
\end{figure}

%Write about calibration and background simulation (like FF, Pedestal and NSB) and especially about trigger (prominently!) 
To understand whether the properties of the single telescopes in an array are defined correctly in simulations, parameters such as the trigger rates, the raw pixel intensity, and the contribution of the pedestal as well as its noise need to be validated for each telescope individually. \\ 
The measured trigger rates are dominated by hadronic showers. Hence, the investigation of the trigger rates is conducted via proton simulations, applying a correction factor for heavier nuclei, which has been validated beforehand with dedicated simulations to be $\approx 1.34 $ for the trigger rates of the entire \hess\ array. 
The raw trigger rates are calculated by folding the effective area with a CR proton spectrum. For high accuracy this work uses the Global Spline Fit spectrum by Hans Dembinsky et al. \cite{global_spline_fit_paper} as it also considers spectral features deviating from a simple power law. The effective area is calculated by taking into account the amount of simulated and triggered showers, simulated area and solid angle.
One has to keep in mind that \hess\ is operated in a hybrid fashion with CT1-5 triggering in stereo and additionally CT5 in mono mode. Hence, for CT1-4 the stereo participation rate and for CT5 the mono rate are the prime quantities to compare \cite{HESS1U}, \cite{ICRC21FCPerf}. \\
In addition to the instrument-specific settings (i.e.\ trigger threshold,  Night Sky Background (NSB) values, PSF$_{80}$ and reflectivity) some general simulation settings (i.e.\ simulated energy range, view cone, photon bunch size\footnote{Simulated Cherenkov photons are stored in so called photon bunches. A bunch size of five is chosen as a compromise between computing time and accuracy.} and atmospheric transmission profile) were adjusted to match real measurements. 
The various adjustments and their effects on the telescope trigger rates are summarised in \autoref{tab:my-table}. \\
The resulting trigger rates can be seen in \autoref{Fig:final_trigger_rates} as a function of zenith angle. For one representative of each of the two \hess\ telescope types, we show both the simulated and real trigger rates. The uncertainty intervals show the investigated systematic uncertainties. For the mirror reflectivity (derived from muon simulations \cite{MuonAlison}) we accepted an uncertainty of $2\,\%$ leading to a systematic uncertainty of $\approx 12\,\%$ in the trigger rates. The systematic uncertainty arising from the choice of the interaction model in the CORSIKA simulation was investigated by repeating the same simulation set with the models QGSJET-2, EPOS, and SYBILL \cite{CorsikaKIT}, \cite{EPOS}. A variation of $\approx 4\,\%$ in the resulting trigger rates was found. In addition, the assumed systematic uncertainty on these rates for the implementation of \hess\ in \simtel\ is $10\,\%$.

\begin{table}[H]
\centering
\begin{tabular}{|c|c|c|c|c|}
\hline
Parameter               & Change CT1-4    & Effect CT1-4 & Change CT5 & Effect CT5\\ \hline
Aerosol level           & $50\,\%$   & $10\,\%$               & $50\,\%$ & $12\,\%$ \\ \hline
Trigger threshold & $27\,\%$   & $41\,\%$               & $4\,\%$ & $6\,\%$\\ \hline
Mirror reflectivity     & $2.4\,\%$ & $15\,\%$               & $8\,\%$& $23\,\%$ \\ \hline
NSB                     & $67\,\%$   & $3- 6\,\%$               & $55\,\%$ & $6\,\%$ \\ \hline
PSF$_{80}$                     & $20\,\%$   & $1\,\%$                & $9\,\%$ & $<1\,\%$ \\ \hline
\end{tabular}
\caption{Change of simulation parameters and their effects on the CT1-4 stereo participation and CT5 mono trigger rates. If the changes and effects differ between CT1-4 the average value is quoted. Presented are the absolute variations and their absolute effects on the trigger rates.}
\label{tab:my-table}
\end{table}

%In addition to the instrument specific settings such as trigger threshold, NSB-values, PSF and reflectivity) some general simulation settings such as simulated energy range, viewcone, photon bunch size and atmospheric transmission profile were adjusted to match real measurements. 

The detector (in \hess: photo-multiplier tubes) output in the cameras is offset by the pedestal.
While measurements are corrected for the mean of this value, its standard deviation ('pedestal width') contributes as noise, mostly caused by diffuse NSB light \cite{HESSCalib2004}. 
Under given observing conditions\footnote{E.g., phase and position of the moon, weather and atmospheric conditions, pointing direction, etc.}, the amount of NSB-photons reaching the cameras depends primarily on the optical brightness of the observed field. 
Here, differences of a factor of 4 or greater are possible \cite{Preuss2002}. 
Therefore, we adjust the pedestal width of the simulations to the median\footnote{Mean is discouraged here, as it would be biased by high-value outliers, e.g.\ from pixels with a star in their Field of View (FoV) that then are automatically switched off.} of observations' pedestal width. 
For this adjustment, the NSB should be set to a reasonable value consistent with calculations that reproduces the required pedestal width in simulations. 
This calculation entails folding the NSB flux at the site of the observatory with the collection area of the telescope and the pixel FoV. 
Provided an NSB value, \simtel\ handles the mirror and quantum efficiencies. 
However, possible reflections of NSB light from the ground back into the camera must be taken into account manually. 
Typical NSB levels are of the order \SI{0.1}{\giga\hertz} for CT1-4 and \SI{0.3}{\giga\hertz} for CT5.
%% Kann bei Platzmangel ersatzlos gelöscht werden:
%In principle, the dark pedestal should be considered as well. That is the pedestal (width) when the cameras don't detect light and are affected by electronic noise only. While this value is supposed to be relatively unchanged compared to the original commissioning of the cameras, it also is irrelevant from which source the pedestal width arises. Therefore, the dark pedestal was neglected in the frame of this work.

\section{Higher level and array validation}
\subsection{Cleaning and background}
\begin{figure}
	\centering
	\subcaptionbox{CT1}%
	[0.45\textwidth]{\centering \includegraphics[width=0.48\textwidth,trim=0 0 0 0, clip]{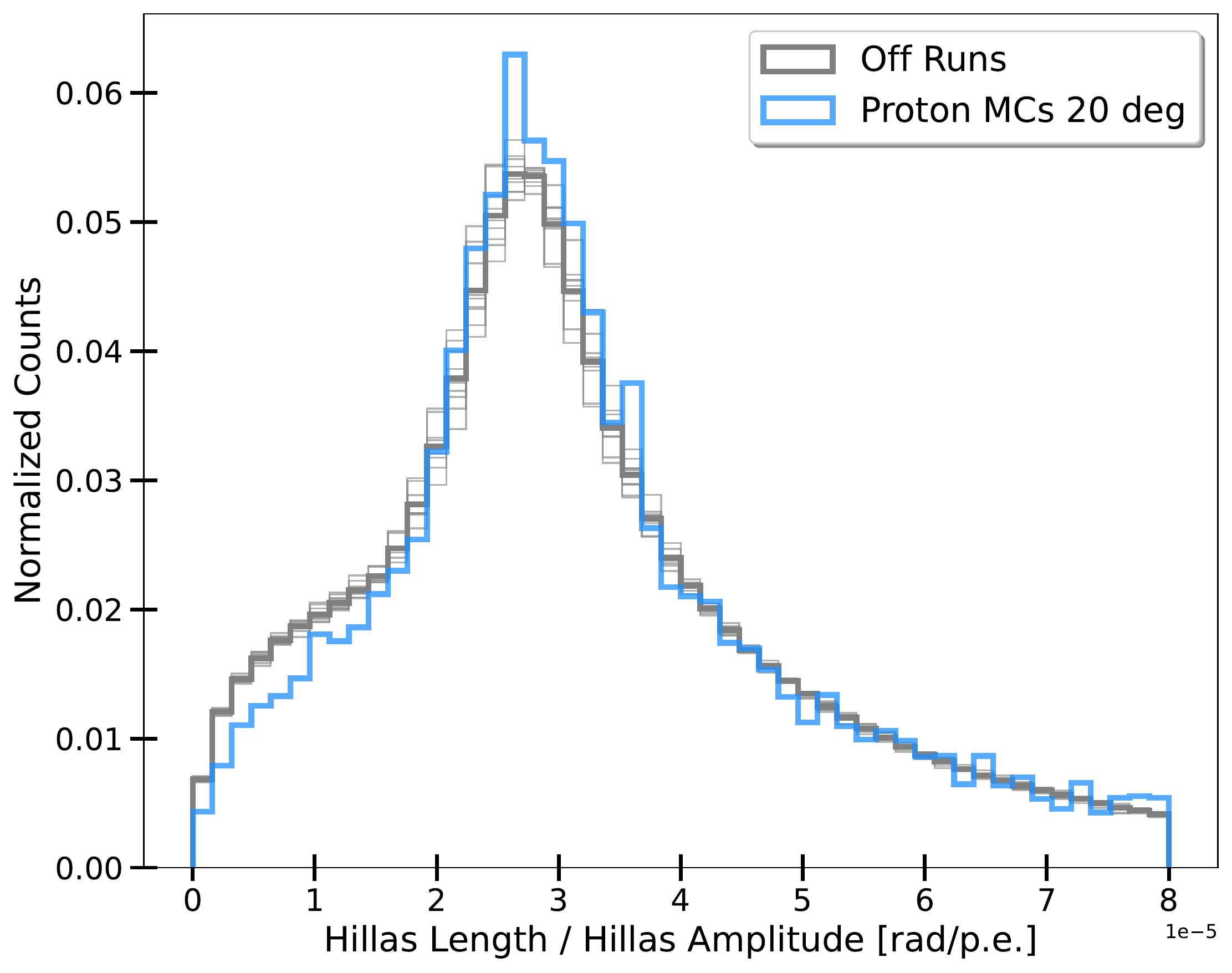}	}
	\quad
	\quad
	\subcaptionbox{CT5}%
	[0.45\textwidth]{\centering \includegraphics[width=0.48\textwidth,trim=0 0 0 0, clip]{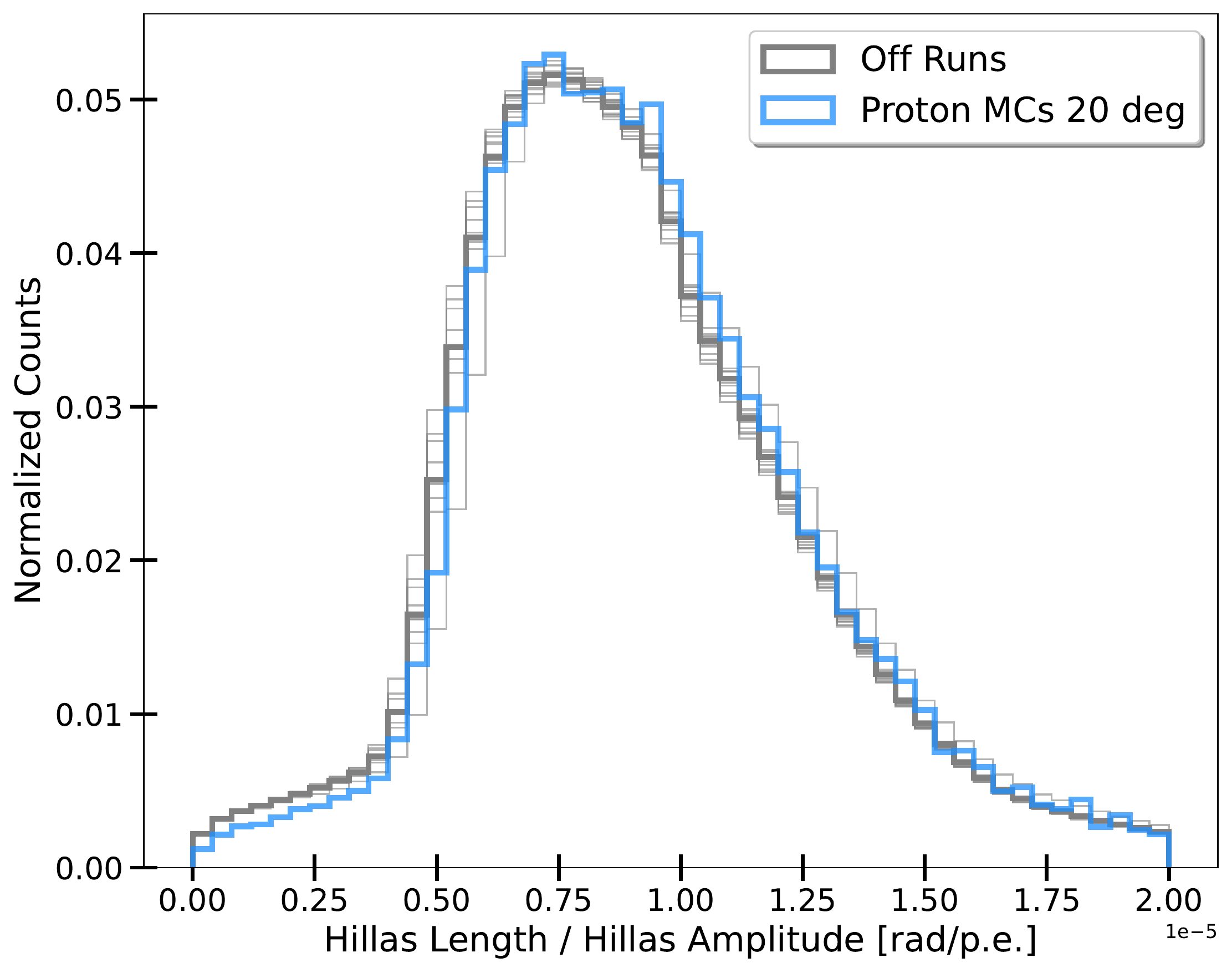}	}

        \caption{The Hillas length divided by the shower intensity of cleaned images for several observation runs compared to proton simulations at 20 deg zenith. The off-run data set consists of 10 individual observation runs with zenith angles between 15 deg and 25 deg. The thick grey line corresponds to the summed distribution. The lighter grey correspond to the individual distributions to show potential run-to-run variability. All distributions were normalised to have a sum of 1.}
	\label{fig:lovers}
\end{figure}

In the analysis chain validated, shower images are cleaned after all calibration steps to avoid any major influence of noise on the image reconstruction. In \hess, cleaning is done using the tail cut method \cite{CrabObs}. From the cleaned images the Hillas parameters \cite{Hillas1985} are derived to describe the image properties.
To test the proper description of the hadronic background the image properties of proton simulations are compared to those of observation runs with no strong gamma source in the FoV (off-runs). The proton simulations were weighted by energy to represent the measured CR proton spectrum \cite{global_spline_fit_paper}. The off-runs were selected to have similar observing conditions as the simulation set, especially in terms of NSB, atmospheric transparency and zenith angle. As an example \autoref{fig:lovers} shows the Hillas length divided by shower intensity distributions for CT1 and CT5. There is a good agreement between data and simulations, that is also reflected in other image parameters. Thereby we conclude that the data-simulation consistency persists also after cleaning. This is crucial as the cleaned images are the basis for the energy \& direction reconstruction in Hillas based analysis algorithms as well as for the gamma-hadron separation.  \\

\subsection{High level validation with the Crab Nebula spectrum}
\begin{figure}[H]
    \centering
    \includegraphics[width = \textwidth]{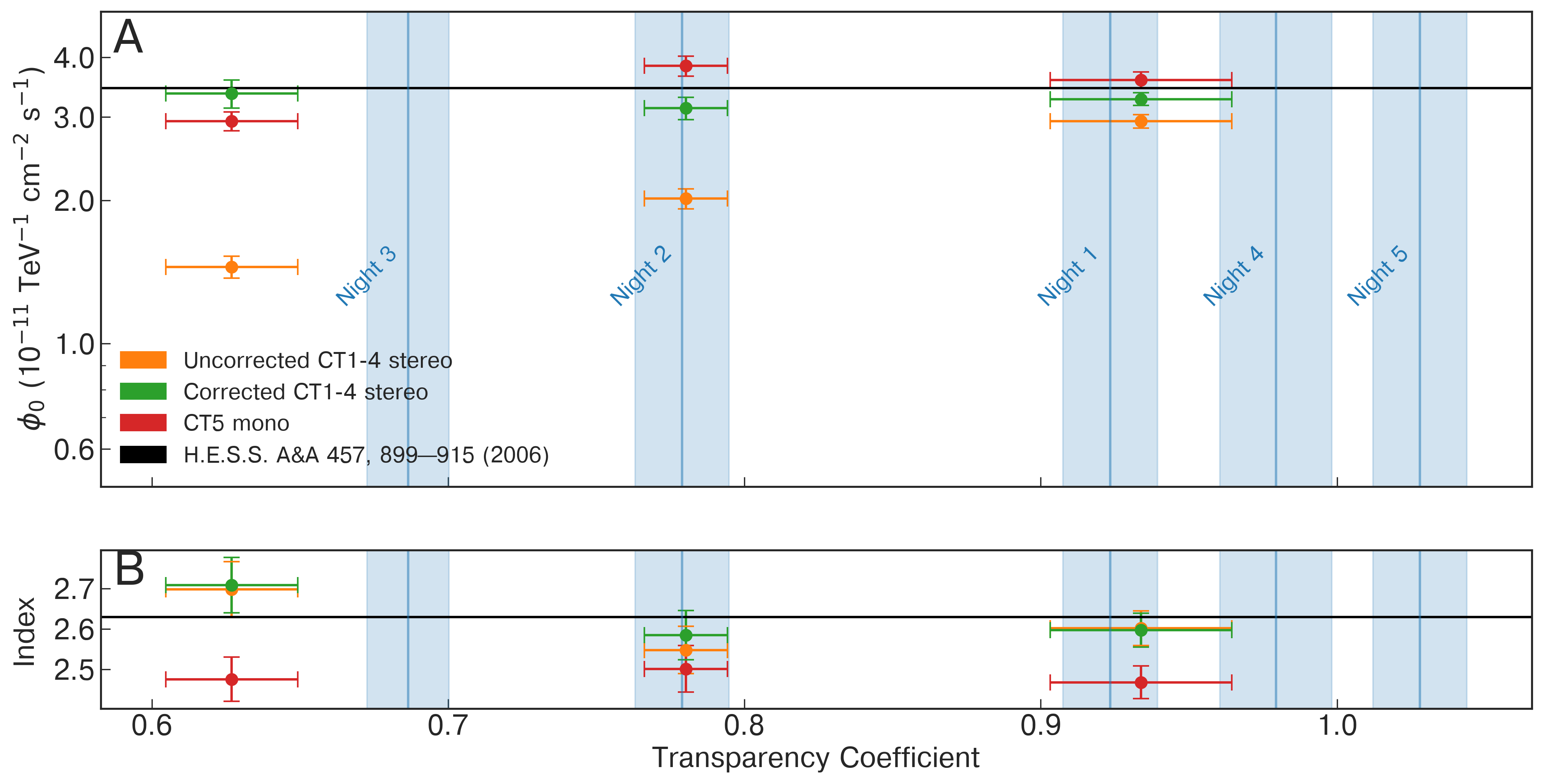}
    \caption{The Crab spectral parameters for different atmospheric conditions as produced for \cite{Hess2022}. The atmospheric conditions are quantified by the Cherenkov transparency coefficient \cite{TransCoeff}. The mono Crab spectra are a result of the work described here and can be compared to corrected stereo results on the same data sets as well as to the Crab spectrum from \cite{CrabObs} with a reference energy of $E_0 = \SI{1}{\tera\eV}$. The blue shaded areas correspond to different atmospheric conditions during the RS Ophiuchi data taking and are not relevant for this work.}
    \label{fig:Crab}
\end{figure}

Energy and direction reconstruction techniques as well as the gamma-hadron separation rely on matching gamma ray simulations. Further, instrument response functions such as effective area, energy dispersion matrix and point spread function are produced from them. To cross check the overall high level performance it is crucial to eventually compare the derived spectrum to a known steady source. For this purpose, the Crab Nebula is selected as the typical standard candle in VHE astrophysics.
Its spectrum was approximated by a power-law model of the form $\upphi_0\left(\frac{E}{E_0}\right)^{-\Gamma}$ to be comparable with the published \hess spectrum \cite{CrabObs}. It can be derived under different atmospheric conditions using simulations including correspondingly different atmospheric transmission profiles. 
This was done already as a cross-check in the published paper on the nova RS Ophiuchi \cite{Hess2022}, for which the monoscopic analysis was based on the simulation configuration derived in this work. The resulting spectral parameters are shown in \autoref{fig:Crab}. 
The monoscopic flux normalisation is consistent with respect to an atmosphere corrected stereo analysis on the same data sets as well as the reference Crab spectrum from \cite{CrabObs} within $15\,\%$. 
The index shows a systematic shift by $\sim 0.2$ that can be explained by the observed hardening of the Crab spectrum towards lower energies. 
This behaviour becomes more important for the mono analysis with its lower energy threshold.

\section{Conclusion and outlook}
%Older phases like old CT5 camera to be validated as well bla bla bla useful for CTA bla bla we're so relevant
We present a procedure to systematically validate all steps of a Monte Carlo simulation configuration for an array of IACTs by comparing simulated to measured properties. It was applied to an analysis chain of the \hess\ experiment with great success. Using the presented method, simulations for future hardware iterations or phases can be adapted and validated with little effort. \\
The presented procedures will be of use for the upcoming CTA observatory as well. When operational, it is possible to quickly assess whether the assumptions made in current MC simulation configurations are correct. Further, a validation of the MC configurations will be necessary e.g.\ when the hardware of CTA will be updated in the future.
Care was taken that the entire framework is well documented for future application. \\ 
Additionally, to the procedure for validating MC configurations, we developed a scheme to assess the impact of short term changes in the atmospheric transmission on the response of atmospheric Cherenkov detectors and to finally correct for them. This scheme is described in detail in a separate work \cite{AtmoHEAD}. It allows to use more of the data obtained by Cherenkov detectors than before and to have a more reliable estimation of primary photon energies.\\

% The combination of these two parts of this work allows to use even single observations to obtain a source's flux and spectrum with high accuracy as can be seen from \autoref{fig:Crab}.

%\begin{thebibliography}{99}
%\bibitem{...}
%....
%
%\end{thebibliography}

%\bibliographystyle{JHEP}
%\bibliography{references}  
\printbibliography

\end{document}